# Highly pure single photon emission from spectrally uniform surface-curvature directed mesa top single quantum dot ordered array

*Jiefei Zhang,[1,2] Swarnabha Chattaraj,[3] Siyuan Lu,[4] and Anupam Madhukar[1,2,3,a]*

[1]Department of Physics and Astronomy, University of Southern California, Los Angeles, California 90089, USA
[2]Mork Family Department of Chemical Engineering and Materials Science, University of Southern California, Los Angeles, California 90089, USA
[3]Ming Hsieh Department of Electrical Engineering, University of Southern California, Los Angeles, California 90089, USA
[4]IBM Thomas J. Watson Research Center, Yorktown Heights, New York, 10598, USA

**Abstract:** Realizing ordered and spectrally uniform single photon source arrays integrable on-chip with light manipulating elements in scalable architecture lies at the core of building monolithic quantum optical circuits (QOCs). We demonstrate here a spatially-ordered 5 × 8 array of surface-curvature driven mesa-top GaAs(001)/InGaAs/GaAs single quantum dots (MTSQDs) that exhibit highly pure (~99% ) single photon emission as deduced from the measured $g^{(2)}(0) < 0.02$ at 9.4K. Polarization-independent and polarization-resolved high resolution photoluminescence (PL) measurements show that these ordered and spectrally uniform QDs have neutral exciton emission with intrinsic linewidth ~ 10 µeV and fine structure splitting < 10 µeV, an important figure of merit for the use of QDs in QOCs. The findings point to the high potential of such MTSQD based single photon source arrays as a promising platform for on-chip scalable integration with light manipulating units (connected resonant cavity, waveguide, beam splitter, etc.) to enable constructing QOCs.

Realizing spatially ordered single photon sources that can be readily integrated with light manipulating elements in a scalable architecture has been a major goal towards realizing on-chip integrated quantum optical circuits[1-3] (QOCs) for applications in quantum communication and quantum information processing (QIP). A significant step towards this goal was recently taken with the demonstration of 5×8 array of a new class of semiconductor single quantum dots that form on the top of laterally confined mesas with unprecedented control on shape and size[4-6]. These mesa-top single quantum dots (MTSQDs) are formed by site-selective size-reducing epitaxy on nanomesas fabricated with specifically chosen edge orientations that induce directed-migration of atoms symmetrically from the sidewalls to the mesa top (Fig. 1(a)) during growth, thus ensuring spatially-selective growth on mesa top[4-6] (Fig.1(b)). The approach is thus dubbed substrate-encoded size-reducing epitaxy (SESRE)[7-9]. The synthesized GaAs(001)/In$_{0.5}$Ga$_{0.5}$As/GaAs MTSQDs (Fig. 1(b)) being reported on here were shown to be efficient single photon emitters at 10K with $g^{(2)}(0)$ ~0.15 and maintain reasonable single photon emission ($g^{(2)}(0)$ ~0.3) up to liquid nitrogen temperature[5,6]. The emission wavelength from every MTSQD in the 5×8 array is shown in Fig.1(c). These MTSQDs are formed with considerable control on size and shape and thus, as grown, exhibit highly uniform PL emission with a standard deviation of ~8 nm, much better than the commonly employed lattice mismatch strain-driven spontaneously formed 3D island quantum dots dubbed self-assembled quantum dots (SAQDs)[5,6].

---

[a] Author to whom correspondence should be addressed. **Electronic mail: madhukar@usc.edu.** Telephone: 1-213-740-4325.

The above noted PL and the encouraging $g^{(2)}(\tau)$ behavior were, however, limited by the instrument resolution of ~300 μeV. The true nature and potential of this new class of epitaxial single QDs (SQDs) was thus not revealed[5]. Strikingly, these studies revealed the presence of pairs of *as-grown* MTSQDs (marked by like-color circles in Fig. 1(c)) with emission within ~300 μeV, a feature that makes this class of SQD arrays particularly attractive and prime candidate for exploring interference and entanglement between photons originating from different but known MTSQDs through their on-chip integration with light manipulating elements (cavity, waveguide, etc.)[5,10,11]. The aim of this letter thus is to report on the true nature of these MTSQDs as revealed by PL, polarization-resolved PL, and $g^{(2)}(\tau)$ studies carried out with the high spetral resolution of ~15 μeV.

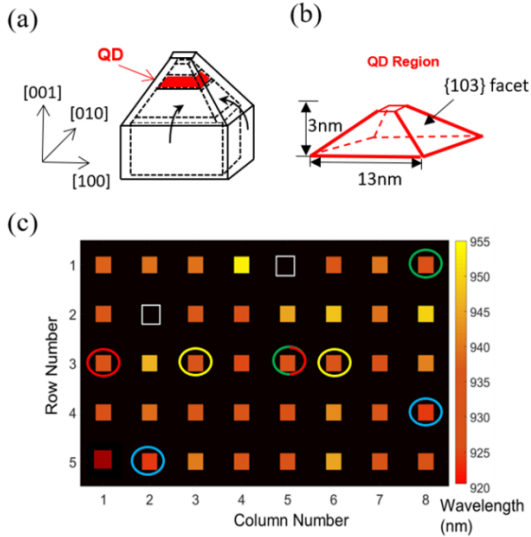

Fig. 1. (a) Schematic of the <100> edge oriented square mesas that induce preferential sidewall-to-mesatop atom migration, thus enabling spatially selective formation of QD (red region) on the (001) top. (b) Schematic of the synthesized mesatop $In_{0.5}Ga_{0.5}As$ SQD with truncated pyramidal shape with {103} side walls. (c) Color-coded plot of the emission wavelength from the 40 MTSQDs in the 5×8 array. Like-color circles mark, strikingly, pairs of *as-grown* MTSQDs with emission within 300 μeV.

The high resolution studies reported here reveal ~99% purity of single photon emission with measured $g^{(2)}(0) < 0.02$. These MTSQDs are found to have neutral exciton emission with linewidths ~10 μeV. Polarization-resolved PL studies reveal the symmetry of the QD confining potential to be $< C_{2v}$. A fine structure splitting (FSS) < 10 μeV is found to accompany the loss of symmetry. The highly pure single photon emission and low FSS highlight the suitability of such QDs as on-chip single photon source arrays that are readily integrable with light manipulating elements (LMEs) (cavity, waveguide, etc.) to realize scalable on-chip quantum optical networks aimed at QIP.

The high resolution PL and polarization-resolved PL data were collected using a μ-PL system that employs a high-resolution (15 μeV) spectrometer (Horiba 1250M) with 1200g/mm grating and a cryogen-free cryostat (Janis CCS-XG-M-204N). A pulsed excitation beam (640 nm 80 MHz diode laser) is focused on a single MTSQD of the array with an excitation spot of diameter of 1.25 μm by a 40X NA0.65 objective. The emitted photons are collected by the same objective, coupled to a single mode optical fiber, spectrally filtered by the spectrometer and detected by a silicon APD (Excelitas SPCM-NIR). Figure 2(a) shows the time- and polarization-integrated PL data from a typical MTSQD's neutral exciton emission collected with a spectral resolution of 15 μeV with pulsed excitation power of 30 nW (2.44 W/cm², 50% of saturation power) at 9.4K. The constant background in Fig. 2(a) is contributed by the APD dark counts. Figure 2(a) shows that the neutral exciton emission is clearly resolved into two peaks $P_1$ (919.108 nm) and $P_2$ (918.891 nm) of unequal intensity separated by 320 μeV. The linewidths (FWHM) of the peak $P_1$ and $P_2$ obtained through fitting Lorentzian shape (red lines in Fig. 2(a)) are found to be ~21 μeV and ~34 μeV, respectively. To reveal the intrinsic linewidths of the peaks, we deconvolute the PL spectrum following a convex optimization method with least square fitting[12]. A Lorentzian with FWHM of 15 μeV is used to represent the independently calibrated instrument response function. The deconvoluted PL data reveal a linewidth of 10 μeV for peak $P_1$ and 24 μeV for peak $P_2$.

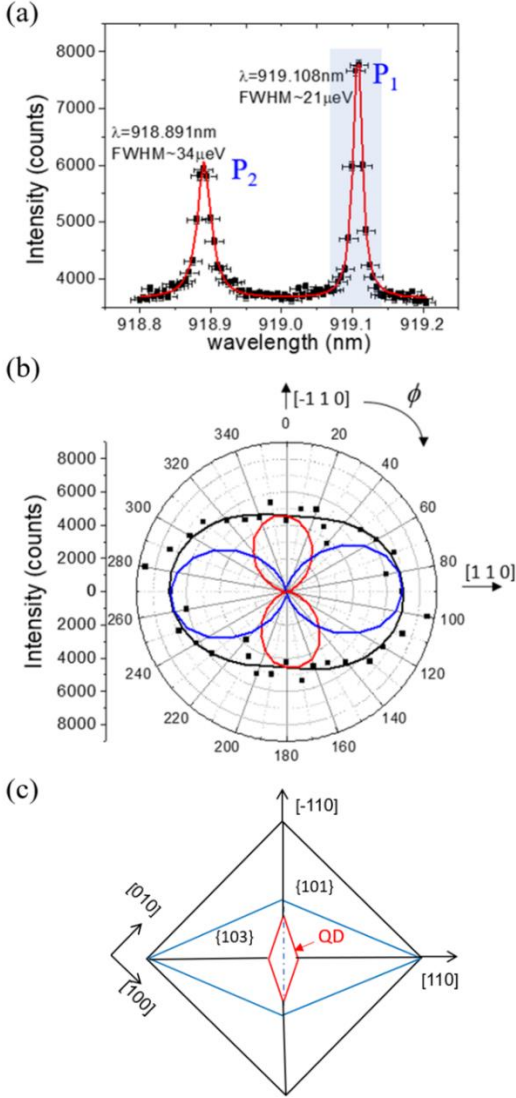

Fig. 2. (a) High resolution PL from a typical MTSQD ((5,2) in the 5 × 8 array) obtained at 9.4 K with 640 nm 80 MHz pulsed laser excitation at excitation power of 30 nW (2.44 W/cm$^2$, 50% of saturation power) and spectral resolution of 15 μeV. (b) The polar plot of the polarization-resolved PL peak intensity (black dots) of peak $P_1$ as a function of the polarizer angle ϕ defined with respect to the [-1 1 0] direction. The black line represents the fitting of the sum of the two linearly polarized FSS states represented by the blue and red curves. (c) Schematic of the SESRE grown mesa top surface profile evolution showing the as-patterned nanomesa edge orientations along <100> directions and the surrounding {103} and {101} facets on the mesa. The QD region evolves to a rhombus shape with base edges (red lines) along <1 -3 0> with (001) top surface, thus lacking the 4-fold symmetry.

To identify the origin of peaks $P_1$ and $P_2$ we recall that a tetrahedrally-bonded III-V semiconductor QD with truncated pyramidal shape (same as the MTSQD shown in Fig. 1(b)) and correspondingly with a confinement potential of $C_{2v}$ symmetry is known[13] to (i) exhibit two finely separated exciton states (bright excitons) with equal oscillator strengths and energy separation typically less than 150 μeV dubbed fine structure splitting (FSS), and (ii) have an optically inactive pair of exciton states (dark excitons) of similar separation. In such a case the QD ground level excitons are constructed of a heavy hole (HH) state characterized by J=3/2, $J_z$=±3/2 and an electron state with S=1/2, $S_z$=±1/2. The excitons formed contain one pair with angular momentum projection $|M|=|J_z-S_z|=1$ coupling to light field and thus dubbed bright excitons and one pair with $|M|=2$ that cannot couple to the light field and dubbed dark exciton.

However, our observation of two peaks with unequal intensity and large energy separation of 320 µeV suggests that the MTSQDs are likely defined by a confinement potential of symmetry < $C_{2v}$ for which the hole state cannot be described as purely HH but as a state with mixed HH and light hole (LH, J=3/2, $J_z=\pm 1/2$) character thereby giving the dark exciton complex a non-zero oscillator strength[13,14]. The peak $P_1$ we anticipate represents the bright exciton complex and hence has two finely separated states with FSS <10 µeV hidden underneath. As discussed next, polarization-resolved PL studies confirm the confinement potential symmetry to be < $C_{2v}$ with the bright exciton (peak $P_1$) comprising two states with a FSS < 10 µeV. Furthermore, analysis of the data sheds light on the degree of heavy hole and light hole mixing.

Figure 2(b) shows the polarization-resolved PL peak intensity data (after APD background counts subtraction) from peak $P_1$ as a function of the polarizer angle ɸ (defined with respect to the crystallographic [-1 1 0] direction) in the *x-y* plane perpendicular to the growth direction [001]. The QD is excited using the same conditions as for the PL data in Fig. 2(a) at 9.4K. A set of PL spectra were collected using a linear polarizer with extinction ratio of $10^4$:1 inserted into the microscope whose angle is adjusted in steps of 10°. Aligning ɸ=0° to the crystallographic [-1 1 0] orientation (with the aid of makers created on the sample) enables direct linkage of the PL to the QD shape as shown in Fig.2 (c). The measured polar pattern[15] is seen to be elliptical with an ellipticity of ~1.65 and the major axis along ~100° direction (10° with respect to the crystallographic [110] direction), the direction along the shorter diagonal of the QD rhombus base (Fig. 2(c)). The observed ellipticity is a clear indication that: (1) the QD has confinement potential symmetry < $C_{2v}$[14,16] consistent with the QD shape being a truncated pyramid with a rhombus base with edges along <1 -3 0> directions lacking the 4-fold symmetry (Fig 2(c)) and the presence of disorder owing to fluctuating indium concentration in the GaAs/In$_{0.5}$Ga$_{0.5}$As/GaAs QD region; (2) the peak $P_1$ arises from the bright exciton complex containing two finely separated peaks [14,16].

As noted above, the ground state exciton in the tetrahedrally-bonded III-V semiconductor based quantum dots of confinement potential symmetry < $C_{2v}$ has mixed $J_z=\pm 3/2$ (HH) and $J_z=\pm 1/2$ (LH) character of the bulk material[16]. In such a case, since the optical transition rules are controlled by only the Bloch part of the wavefunctions while the envelop function part merely contributes to a constant factor, the transition dipole moments of the two FSS states of the bright exciton can be denoted as $<u_E^+|-e\vec{r}|u_H^+> + <u_E^-|-e\vec{r}|u_H^->$ and $<u_E^+|-e\vec{r}|u_H^+> - <u_E^-|-e\vec{r}|u_H^->$, where $|u_E^{\pm}>$ and $|u_H^{\pm}>$ represent, respectively, the Bloch parts of the wavefunction of the electron and the mixed hole states. The mixed HH and LH character of the QD hole states involved in the transitions with the electron states $|u_E^{\pm}> = |\frac{1}{2}, \pm\frac{1}{2}>$ is now represented by the following[16]: $|u_H^{\pm}> = \sqrt{1-\beta^2-\gamma^2}|\frac{3}{2},\pm\frac{3}{2}> + \beta e^{\pm 2i\theta}|\frac{3}{2},\mp\frac{1}{2}> \pm \gamma e^{\pm 2i\varphi}|\frac{3}{2},\pm\frac{1}{2}>$. Here $|u_H^{\pm}>$ is described using the Luttinger-Kohn basis with β and γ representing the amplitude, and θ and φ being the phase of mixing between the $|\frac{3}{2},\pm\frac{3}{2}>$ hole state with the $|\frac{3}{2},\pm\frac{1}{2}>$ and $|\frac{3}{2},\mp\frac{1}{2}>$ hole states, respectively. One may further show that the photons from such two FSS states approximated as point transition-dipoles with the above-mentioned dipole moments will produce polar patterns with ellipticity $e = \frac{1-\frac{2}{3}\beta^2-\gamma^2+2\beta\sqrt{\frac{1-\beta^2}{3}}}{1-\frac{2}{3}\beta^2-\gamma^2-2\beta\sqrt{\frac{1-\beta^2}{3}}}$ when the dielectric effect of the surrounding medium and the effect of the measurement geometry are neglected. The greater than one ellipticity observed in Figure 2(b) thus qualitatively indicates that the MTSQD has a nonzero β and has a confinement potential of less than $C_{2v}$ symmetry, unlike QDs with $C_{2v}$ that have a circular polar pattern[14,16]. To quantify the degree of intrinsic mixing of the HH and LH, we calculate the integrated photon flux within the collection cone of the objective lens from the two FSS states as a function of polarization. The calculation employs a finite element method and assumes that the two states emit as point dipoles (as discussed above) embedded in GaAs nanomesa of size and shape obtained from SEM images (not shown but similar to Fig. 1(a)). The parameters β, γ, θ, φ, are used as fitting parameters to compare the calculated results with the measured data shown in Fig.2(b). We find that the measured data can be explained by the combined effect of two linearly polarized FSS states shown as the black curve in Fig 2(b) with the two FSS states polarized primarily along [110] (blue curve in Fig. 2(b) representing the dipole element $<u_E^+|-e\vec{r}|u_H^+> + <u_E^-|-e\vec{r}|u_H^->$) and [1-10] (red curve in Fig. 2(b) representing the dipole element $<u_E^+|-e\vec{r}|u_H^+> - <u_E^-|-e\vec{r}|u_H^->$) directions but with different amplitudes due to the mixing of HH-LH manifold represented by |β|=0.25±0.02. Limited by the measurement geometry in the *x-y* plane, the parameter γ cannot be obtained from the data and the fitting.

The polarization dependent PL, we thus conclude, indicates that the $P_1$ peak with linewidth ~10 μeV is likely from the intrinsic bright neutral exciton comprising two non-degenerate FSS states. Their linewidths and splitting (the FSS) is thus less than 10 μeV (i. e. below our resolution limit). This FSS is comparable to the best reported for other types of QDs such as the ordered QDs in recesses[17], the typical 3D island based SAQDs[1-3,18], and the well explored nanowire QDs[19,20]. Such a low FSS in the MTSQDs in spatially regular arrays is a highly encouraging figure of merit for their use in QOCs. Thus we next present measurements of the two photon emission correlation function for photons emitted from the bright exciton peak $P_1$ to examine the intrinsic purity of the single photon emission from the MTSQD.

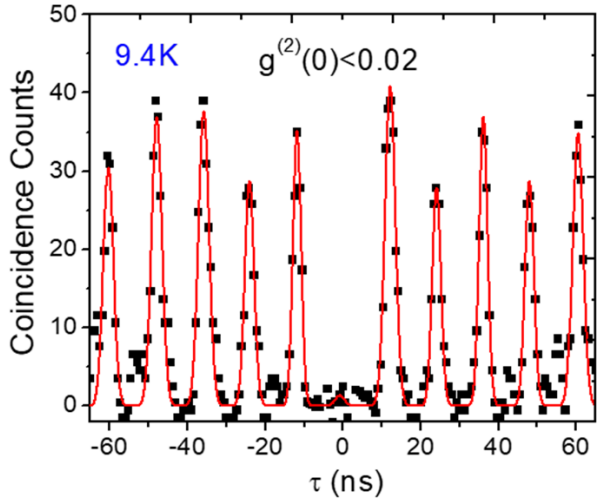

Fig. 3 Coincidence count histogram of MTSQD (5,2) at 9.4K with background contributed from APDs dark counts subtracted. The obtained $g^{(2)}(0)$ value is less than 0.02 as extracted from the measured data with the detector dark counts subtracted. The red line shows the fitting of the measured data confirming the ultra-low $g^{(2)}(0)$

A standard Hanbury Brown – Twiss (HBT) instrumentation is employed for measurements of two photon coincidence counts. The emitted photons from MTSQD bright exciton (peak $P_1$, Fig. 2(a)), collected at 9.4K under the same excitation as described before using an acceptance window of ~70 μeV (indicated by the blue shade in Fig. 2(a)), are directed towards the HBT setup with its two detectors for measuring coincident counts. Figure 3 shows the measured histogram of the coincidence counts (black dots) as a function of τ, the time difference between the detection events at the two APD detectors. The background count contributed by the dark counts from the two Si APDs has been subtracted in the plot. The $g^{(2)}(0)$ is obtained from calculating the ratio of the τ=0 peak area to the average of the other peaks. The $g^{(2)}(0)$ is found to be almost zero with a upper bound of 0.02, indicating that the single photon emission purity, $\sqrt{1-g^{(2)}(0)}$, of MTSQDs is around 99%. The ultra-low $g^{(2)}(0)$ is also confirmed from the fitting of the measured data shown as red lines in Fig. 3 with the near zero peak at τ=0. The revealed highly pure (around 99%) single photon emission from MTSQD is comparable to other SQD based best SPSs[1-3,17,19-25] reported in the literature, such as the QD in recesses[17,21], the SAQDs[1-3,23], and the nanowire SQDs[19,20,24,25]. As reported previously[5,6], these MTSDQ array SQDs can provide single photon emission even at 77K with single photon emission purity ~80% ($g^{(2)}(0)$~0.3). The intrinsic purity and the robustness of single photon emission at elevated temperature from such spatially ordered and *as-grown* highly spectrally uniform SQDs suggests that MTSQDs are a highly promising candidate for single photon sources for on-chip integration with LMEs such as resonant cavity, waveguide, etc. for realizing on-chip QOCs.

In summary, we have demonstrated that the spectrally uniform InGaAs MTSQD array containing as-grown pairs of QDs emitting within 300 μeV synthesized using SESRE approach have sharp exciton emissions with intrinsic linewidth of 10 μeV and FSS < 10 μeV, a figure of great importance for QD potential use in QOCs. More importantly, the ordered uniform MTSQDs can emitted highly pure single photons with purity around 99% as deduced from the measured $g^{(2)}(0)$ < 0.02 at 9.4K. The purity of single photon emission from this new class of ordered and spectrally uniform QDs being comparable to the best reported for other classes of SQDs not necessarily in ordered arrays in the literature[1-3, 22, 23] makes this new class of QDs a promising candidate for single photon source to be on-chip integrated with LMEs to realize optical circuits.

We close noting that with the overgrowth of a planarizing layer, similar to QD in recesses[26,27], the MTSQDs can be embedded in GaAs layer with flat top surface, enabling ready integration with lithographically carved light manipulating elements as discussed in references 5 and 11. The many favorable properties of the MTSQDs provide strong incentive to further explore the paradigm of using these SQD arrays to construct on-chip QOCs by integrating them with the typically well explored 2D photonic crystal based light manipulating elements[23, 28-30] or, alternatively, using the newer approach of exploiting a single Mie resonance of co-designed network built of subwavelength sized dielectric building blocks (DBBs)[5,11] to provide the simultaneously needed light manipulating functions of enhancing photon emission rate, directing photon emission, guiding, beam splitting and combing on-chip. Further studies of MTSQDs examining their coherence time and indistinguishability of emitted photon as well as on integrating these QD with DBBs based light manipulating elements are underway.

## Acknowledgements

Army Research Office (ARO) (W911NF-15-1-0298); Air Force Office of Scientific Research (AFOSR) (FA9550-17-01-0353 and FA9550-10-01-0066).